\documentclass[aps,prl,twocolumn,superscriptaddress]{revtex4}
\usepackage{graphicx}

\begin{document}


\title{Evidence for radial flow of thermal dileptons in high-energy nuclear collisions}




\author{R.~Arnaldi}
\affiliation{Universit\`a di Torino and INFN,~Italy}
\author{K.~Banicz}
\affiliation{CERN, 1211 Geneva 23, Switzerland}
\affiliation{Physikalisches~Institut~der~Universit\"{a}t Heidelberg,~Germany}
\author{J.~Castor}
\affiliation{LPC, Universit\'e Blaise Pascal and CNRS-IN2P3, Clermont-Ferrand, France}
\author{B.~Chaurand}
\affiliation{LLR, Ecole Polytechnique and CNRS-IN2P3, Palaiseau, France}
\author{C.~Cical\`o}
\affiliation{Universit\`a di Cagliari and INFN, Cagliari, Italy}
\author{A.~Colla}
\affiliation{Universit\`a di Torino and INFN,~Italy}
\author{P.~Cortese}
\affiliation{Universit\`a di Torino and INFN,~Italy}
\author{S.~Damjanovic}
\affiliation{CERN, 1211 Geneva 23, Switzerland}
\affiliation{Physikalisches~Institut~der~Universit\"{a}t Heidelberg,~Germany}
\author{A.~David}
\affiliation{CERN, 1211 Geneva 23, Switzerland}
\affiliation{Instituto Superior T\'ecnico, Lisbon, Portugal}
\author{A.~de~Falco}
\affiliation{Universit\`a di Cagliari and INFN, Cagliari, Italy}
\author{A.~Devaux}
\affiliation{LPC, Universit\'e Blaise Pascal and CNRS-IN2P3, Clermont-Ferrand, France}
\author{L.~Ducroux}
\affiliation{IPN-Lyon, Universit\'e Claude Bernard Lyon-I and CNRS-IN2P3, Lyon, France}
\author{H.~En'yo}
\affiliation{RIKEN, Wako, Saitama, Japan}
\author{J.~Fargeix}
\affiliation{LPC, Universit\'e Blaise Pascal and CNRS-IN2P3, Clermont-Ferrand, France}
\author{A.~Ferretti}
\affiliation{Universit\`a di Torino and INFN,~Italy}
\author{M.~Floris}
\affiliation{Universit\`a di Cagliari and INFN, Cagliari, Italy}
\author{A.~F\"orster}
\affiliation{CERN, 1211 Geneva 23, Switzerland}
\author{P.~Force}
\affiliation{LPC, Universit\'e Blaise Pascal and CNRS-IN2P3, Clermont-Ferrand, France}
\author{N.~Guettet}
\affiliation{CERN, 1211 Geneva 23, Switzerland}
\affiliation{LPC, Universit\'e Blaise Pascal and CNRS-IN2P3, Clermont-Ferrand, France}
\author{A.~Guichard}
\affiliation{IPN-Lyon, Universit\'e Claude Bernard Lyon-I and CNRS-IN2P3, Lyon, France}
\author{H.~Gulkanian}
\affiliation{YerPhI, Yerevan Physics Institute, Yerevan, Armenia}
\author{J.~M.~Heuser}
\affiliation{RIKEN, Wako, Saitama, Japan}
\author{M.~Keil}
\affiliation{CERN, 1211 Geneva 23, Switzerland}
\affiliation{Instituto Superior T\'ecnico, Lisbon, Portugal}
\author{L.~Kluberg}
\affiliation{CERN, 1211 Geneva 23, Switzerland}
\affiliation{LLR, Ecole Polytechnique and CNRS-IN2P3, Palaiseau, France}
\author{C.~Louren\c{c}o}
\affiliation{CERN, 1211 Geneva 23, Switzerland}
\author{J.~Lozano}
\affiliation{Instituto Superior T\'ecnico, Lisbon, Portugal}
\author{F.~Manso}
\affiliation{LPC, Universit\'e Blaise Pascal and CNRS-IN2P3, Clermont-Ferrand, France}
\author{P.~Martins}
\affiliation{CERN, 1211 Geneva 23, Switzerland}
\affiliation{Instituto Superior T\'ecnico, Lisbon, Portugal}
\author{A.~Masoni}
\affiliation{Universit\`a di Cagliari and INFN, Cagliari, Italy}
\author{A.~Neves}
\affiliation{Instituto Superior T\'ecnico, Lisbon, Portugal}
\author{H.~Ohnishi}
\affiliation{RIKEN, Wako, Saitama, Japan}
\author{C.~Oppedisano}
\affiliation{Universit\`a di Torino and INFN,~Italy}
\author{P.~Parracho}
\affiliation{CERN, 1211 Geneva 23, Switzerland}
\author{P.~Pillot}
\affiliation{IPN-Lyon, Universit\'e Claude Bernard Lyon-I and CNRS-IN2P3, Lyon, France}
\author{T.~Poghosyan}
\affiliation{YerPhI, Yerevan Physics Institute, Yerevan, Armenia}
\author{G.~Puddu}
\affiliation{Universit\`a di Cagliari and INFN, Cagliari, Italy}
\author{E.~Radermacher}
\affiliation{CERN, 1211 Geneva 23, Switzerland}
\author{P.~Ramalhete}
\affiliation{CERN, 1211 Geneva 23, Switzerland}
\author{P.~Rosinsky}
\affiliation{CERN, 1211 Geneva 23, Switzerland}
\author{E.~Scomparin}
\affiliation{Universit\`a di Torino and INFN,~Italy}
\author{J.~Seixas}
\affiliation{Instituto Superior T\'ecnico, Lisbon, Portugal}
\author{S.~Serci}
\affiliation{Universit\`a di Cagliari and INFN, Cagliari, Italy}
\author{R.~Shahoyan}
\affiliation{CERN, 1211 Geneva 23, Switzerland}
\affiliation{Instituto Superior T\'ecnico, Lisbon, Portugal}
\author{P.~Sonderegger}
\affiliation{Instituto Superior T\'ecnico, Lisbon, Portugal}
\author{H.~J.~Specht}
\affiliation{Physikalisches~Institut~der~Universit\"{a}t Heidelberg,~Germany}
\author{R.~Tieulent}
\affiliation{IPN-Lyon, Universit\'e Claude Bernard Lyon-I and CNRS-IN2P3, Lyon, France}
\author{G.~Usai}
\affiliation{Universit\`a di Cagliari and INFN, Cagliari, Italy}
\author{R.~Veenhof}
\affiliation{CERN, 1211 Geneva 23, Switzerland}
\author{H.~K.~W\"ohri}
\affiliation{Universit\`a di Cagliari and INFN, Cagliari, Italy}
\affiliation{Instituto Superior T\'ecnico, Lisbon, Portugal}

\collaboration{NA60 Collaboration}\noaffiliation



\date{\today}

\begin{abstract}

The NA60 experiment at the CERN SPS has studied low-mass dimuon
production in 158 AGeV In-In collisions. An excess of pairs above the
known meson decays has been reported before. We now present precision
results on the associated transverse momentum spectra. The slope
parameter $T_\mathrm{eff}$ extracted from the spectra rises with
dimuon mass up to the $\rho$, followed by a sudden decline
above. While the initial rise is consistent with the expectations for
radial flow of a hadronic decay source, the decline signals a
transition to an emission source with much smaller flow. This may well
represent the first direct evidence for thermal radiation of partonic
origin in nuclear collisions.

\end{abstract}

\maketitle

Among the observables used for the diagnostics of the hot and dense
fireball formed in high-energy nuclear collisions, lepton pairs are
particularly attractive. In contrast to hadrons, they directly probe
the entire space-time evolution of the fireball and freely escape,
undisturbed by final-state interactions. In the invariant mass region
$\leq$1 GeV, thermal dilepton production is largely mediated by the
broad vector meson $\rho$(770). Due to its strong coupling to the
$\pi\pi$ channel and its short lifetime of only 1.3 fm/c, its
in-medium properties like mass and width have long been considered as
the prime signature for the restoration of chiral symmetry, associated
with the QCD phase transition from hadronic to partonic
matter~\cite{Pisarski:mq,Brown:2001nh,Rapp:1999ej}. In the mass region
$>$1 GeV, thermal dileptons may be produced in either the early
partonic or the late hadronic phase of the fireball, based here on
hadronic processes other than $\pi\pi$ annihilation.

In contrast to real photons, virtual photons decaying into lepton
pairs can be characterized by two variables, mass $M$ and transverse
momentum $p_{T}$. Historically, the interest has largely focused on
mass, including most recently the first measurement of the space-time
averaged $\rho$ spectral function in nuclear
collisions~\cite{Arnaldi:2006jq}. Transverse momentum, on the other
hand, contains not only contributions from the spectral function, but
encodes the key properties of the expanding fireball, temperature and
in particular transverse (radial) flow. In the description of hadron
$p_{T}$ spectra, the study of collective flow has contributed
decisively to the present understanding of the fireball dynamics in
nuclear collisions~\cite{Schnedermann:1993ws,Heinz:2004qz}. However,
while hadrons always receive the full asymptotic flow reached at the
moment of decoupling from the flowing medium, lepton pairs are
continuously emitted during the evolution, sensing small flow and high
temperature at early times, and increasingly larger flow and smaller
temperatures at later times. The resulting space-time folding over the
temperature-flow history can be turned into a diagnostic tool: the
measurement of $p_{T}$ spectra of lepton pairs potentially offers
access to their emission region and may thereby differentiate between
a hadronic and a partonic nature of the emitting
source~\cite{Renk:2007pr}.

Experimentally, dilepton $p_{T}$ spectra associated with direct
radiation in the mass region $\leq$1 GeV were previously only
investigated by the CERES/NA45 experiment at the CERN
SPS~\cite{Agakichiev:1997au}, but with low statistics. In this Letter,
we present the first precise data on such spectra, obtained by the
NA60 experiment at the CERN SPS for 158 AGeV In-In collisions. Mass
spectra in different $p_{T}$ windows~\cite{Damjanovic:2006bd},
uncorrected for acceptance, and preliminary results on
acceptance-corrected $p_{T}$ spectra were presented
before~\cite{Damjanovic:2007qm,GU_JS:2006qm}.

\begin{figure}[t!]
\begin{center}
\includegraphics*[width=6.8cm, height=7.3cm,clip=, bb= 0 12 560 665]{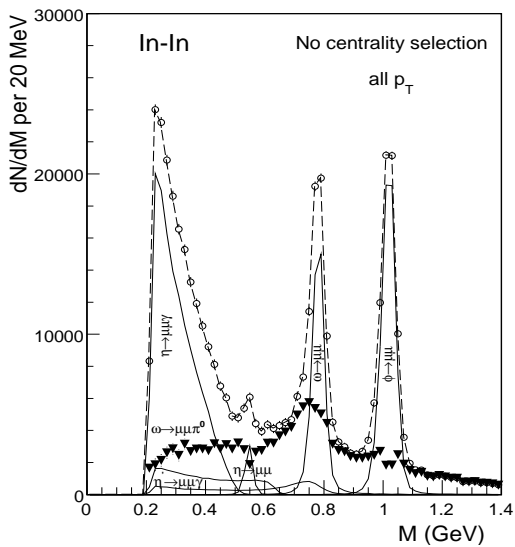}
\vspace*{-0.3cm}
\caption{Isolation of an excess above the electromagnetic decays of
neutral mesons (see text). Total data (open circles), individual
cocktail sources (solid), difference data (thick triangles), sum of
cocktail sources and difference data (dashed). Open charm still
contained.}
   \label{fig1}
\end{center}
\vspace*{-0.8cm}
\end{figure}

Details of the NA60 apparatus are contained
in~\cite{Banicz:2005nz,Gluca:2005}. The different analysis steps
follow the same sequence as described in our previous
Letter~\cite{Arnaldi:2006jq}, including the critical assessment of the
combinatorial background from $\pi$ and $K$ decays by a mixed-event
technique~\cite{Shahoyan:2006eb}. The centrality-integrated net mass
spectrum of the final data sample used here is shown in
Fig.~\ref{fig1}. It contains about 430\,000 dimuons in the mass range
$\leq$ 1.4 GeV. The narrow vector mesons $\omega$ and $\phi$ are
completely resolved; the mass resolution at the $\omega$ is 20
MeV. Fig.~\ref{fig1} also contains the expected contributions from the
electromagnetic decays of neutral mesons, i.e. the 2-body decays of
the $\eta$, $\omega$ and $\phi$ resonances and the Dalitz decays of
the $\eta$, $\eta^{'}$ and $\omega$. The peripheral data can be
described by the sum of these ``cocktail'' contributions plus the
$\rho$~\cite{Arnaldi:2006jq,Damjanovic:2006bd}. This is not possible
for the total data as plotted in Fig.~\ref{fig1}, due to the existence
of a strong excess of pairs.  To isolate this excess with {\it a
priori unknown characteristics} without any fits, the cocktail of the
decay sources is subtracted from the data using {\it local} criteria
which are solely based on the measured mass distribution itself. The
$\rho$ is not subtracted. The procedure is illustrated in
Fig.~\ref{fig1}. Each centrality window is treated separately. The
space of $p_{T}$ is subdivided in the range 0$\leq$$p_{T}$$\leq$2 GeV
into 10 bins of equal width, and each bin is also treated
separately. The yields of the narrow vector mesons $\omega$ and $\phi$
are fixed so as to get, after subtraction, a {\it smooth} underlying
continuum, leading to an accuracy of about 2\% for the $\phi$ and
3-4\% for the $\omega$; at very low $p_{T}$, even an error of 20\% on
the $\omega$ would still translate to an error of only 5\% of the
excess in the mass region 0.6$<$$M$$<$0.9 GeV. The yield of the $\eta$
relative to the $\omega$ and $\phi$, relevant only for masses
$\leq$0.4 GeV, is fixed from the data at p$_{T}$$>$1.0 GeV, profiting
from the very high sensitivity of the spectral shape of the Dalitz
decay to any underlying admixture from other sources. The yield at
lower $p_{T}$, subject to lower statistics due to acceptance, is
determined using the NA60 hadron-decay generator
GENESIS~\cite{genesis:2003} tuned to the cocktail data. The estimated
uncertainty on the $\eta$ subtraction is $\sim$3\%, transforming to a
systematic error of $\sim$20\% on the excess in the mass region
0.2$<$$M$$<$0.4 GeV. Having fixed the three main sources, the $\eta$
two-body and $\omega$ Dalitz decays are then bound as well; the ratio
$\eta^{'}$/$\eta$ is assumed to be 0.12~\cite{genesis:2003}.  Open
charm, measured to be 0.30$\pm$0.06 of the total yield in the mass
interval 1.2$<$$M$$<$1.4 GeV by NA60~\cite{Shahoyan:2006qm}, is
subtracted throughout (but not yet in Fig.~\ref{fig1}), with the
spectral shape in $M$ and $p_{T}$ as described by
Pythia~\cite{Shahoyan:2006qm}; the resulting uncertainty of the excess
is 8\% in this region and $\leq$1\% everywhere else. After subtraction
of the meson decays and charm, the remaining sample contains
$\sim$150\,000 dimuons. The subtracted data for the $\eta$, $\omega$
and $\phi$ themselves are subject to the same further steps as the
excess data and are used later for comparison.

In the last step, the data are corrected for the acceptance of the
NA60 apparatus and for the centrality-dependent pair reconstruction
efficiencies.  
\begin{figure}[h!]  
\begin{center} 
\includegraphics*[width=6.2cm, height=5.2cm, clip= 0 0 548 538]{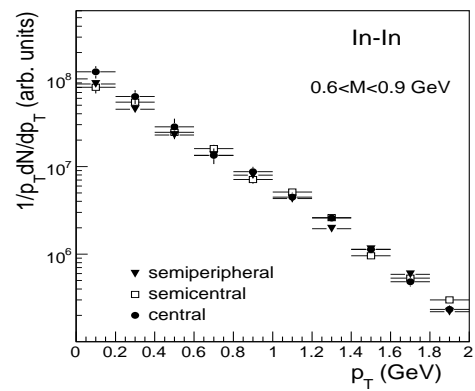} 
\vspace*{-0.3cm} 
\caption{Transverse momentum
spectra of the rho-like mass region for three centrality windows,
arbitrarily normalized. The errors are purely statistical, for
systematic errors see text.}  
\label{fig2} 
\end{center}
\vspace*{-0.4cm} 
\end{figure} 
The acceptance shows strong variations with mass and
$p_{T}$~\cite{Damjanovic:2006bd,Damjanovic:2007qm}, but is understood
on the level of $<$10\%, mainly based on a detailed study of the
peripheral data for the particle ratios $\eta/\omega$ and
$\phi/\omega$~\cite{Arnaldi:2006jq,Damjanovic:2006bd}. In principle,
the acceptance correction requires a 3-dimensional grid in
$M$-$p_{T}$-$y$ space. To avoid large statistical errors in
low-acceptance bins, the correction is performed in 2-dimensional
$M$-$p_{T}$ space (with 0.1 GeV bins in mass and 0.2 GeV bins in
$p_{T}$), using the measured rapidity distribution of the excess as an
input~\cite{Damjanovic:2007qm}. The latter is determined with an
acceptance correction in $y$ found, in an iterative way, from MC
simulations matched to the data in $M$ and $p_{T}$. Separately for
each centrality window, an overlay MC method is used to include the
effects of pair reconstruction efficiencies. The resulting values vary
from mostly $\geq$0.9 down to about 0.7 at low mass and low transverse
momentum for the highest centrality window. All sources are simulated
with a uniform distribution in cos$\theta_{CS}$, where $\theta_{CS}$
is the polar angle of the muons in the Collins-Soper frame, consistent
with (yet unpublished) NA60 data for the $\omega$, the $\phi$ and the
excess.

\begin{figure}[t!]
\begin{center}
\includegraphics*[width=0.34\textwidth, clip= 0 23 449 715]{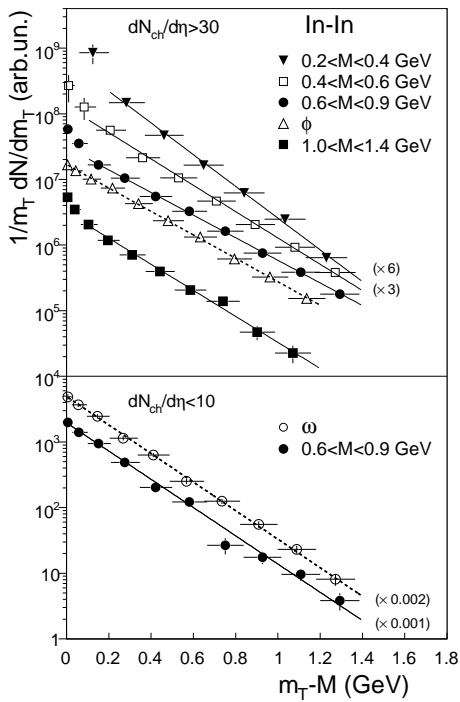}
\vspace*{-0.3cm}
   \caption{Upper: Transverse mass spectra of the excess for four
   mass windows summed over centralities (excluding the peripheral
   bin), in comparison to the $\phi$. Lower: Transverse mass spectrum
   of the rho-like mass region (without $\omega$) for very peripheral
   events, in comparison to the $\omega$. The errors are purely statistical,
   for systematic errors see text.}
   \label{fig3}
\end{center}
\vspace*{-0.7cm}
\end{figure}

\begin{figure}[t!]
\begin{center}
\includegraphics*[width=0.43\textwidth, clip= 0 2 564 537]{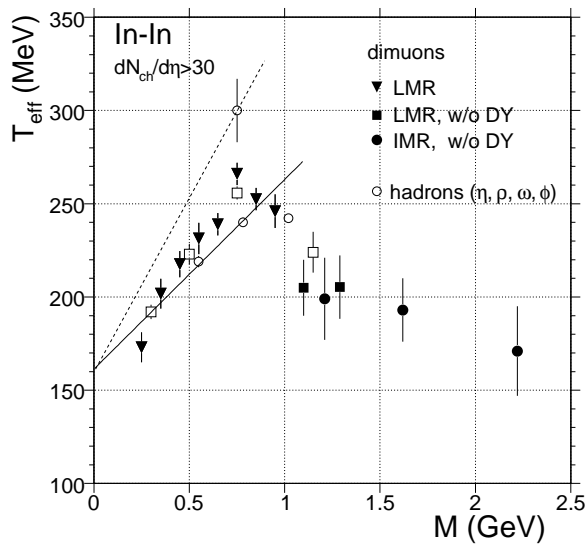}
\vspace*{-0.3cm}
   \caption{Inverse slope parameter $T_\mathrm{eff}$ vs. dimuon mass
   $M$ for $dN_{ch}/d\eta$$>$30. The open squares correspond to the
   fit lines in Fig.~\ref{fig3}. Open charm is subtracted
   throughout. For explanation of the inserted symbols and the errors
   see text.}
   \label{fig4}
\end{center}
\vspace*{-0.7cm}
\end{figure}

Results on acceptance-corrected $p_{T}$-spectra for the mass window
0.6$\leq$$M$$\leq$0.9 GeV and the three upper centralities are shown
in Fig.~\ref{fig2}; equivalent data for other mass windows are
contained in~\cite{Damjanovic:2007qm}. Systematic errors, not
contained in Fig.~\ref{fig2}, mainly arise from the uncertainties of
the combinatorial-background and fake-matches subtraction and range
from 10\% to 30\% for semiperipheral up to central collisions at low
$p_{T}$, decreasing rapidly to a level of only a few \% at higher
$p_{T}$. All other uncertainties are discussed below, in connection
with Fig.~\ref{fig4}. The data of the three centrality windows agree
within their errors; this also holds for the other mass
windows~\cite{Damjanovic:2007qm}.

Fig.~\ref{fig3} (upper) displays the centrality-integrated data
vs. transverse mass $m_{T}$, where $m_{T}=(p_{T}^{2}+M^{2})^{1/2}$,
for four mass windows; the $\phi$ is included for comparison. The
systematic errors due to the background subtraction are, at low
$m_{T}$, 30\%, 25\%, 15\% and 15\% in the four windows, respectively,
falling again rapidly to a level of a few \% at high $m_{T}$; for the
$\phi$, they are smaller by a factor of 5. At very low $m_{T}$, a
steepening is observed in all four mass windows, reminiscent of pion
spectra and opposite to the expectation for radial flow at masses
above the pion mass. This increased rise cannot be due to unsubtracted
background which increases by a factor of 3 vs. centrality, while the
data are independent of it (see above). It cannot be due to
ill-understood acceptance either, since the $\phi$, placed just
between the two upper mass windows, does not show it. While some rise
still persists in the peripheral bin, it finally disappears for very
peripheral collisions with 4$<$$dN_{ch}/d\eta$$<$10 as shown in
Fig.~\ref{fig3} (lower); the $\omega$ does not show it either. The
lines in Fig.~\ref{fig3} are fits with the function 1/$m_{T}$
$dN/dm_{T}$ $\propto $ $exp(-m_{T}/T_\mathrm{eff})$, where the
effective temperature parameter $T_\mathrm{eff}$ characterizes the
slope of the distributions. For the excess data in Fig.~\ref{fig3}
(upper), the fits are restricted to the range 0.4$<$$p_{T}$$<$1.8 GeV
(roughly 0.1$<$$(m_{T} - M)$$<$1.2 GeV), to exclude the increased rise
at low $m_{T}$; for all other spectra, the fits start at zero.

The extracted values of $T_\mathrm{eff}$ vs. pair mass are summarized
in Fig.~\ref{fig4} (open squares), supplemented by a set of further
fit values from narrow slices in $M$ (closed triangles).  Preliminary
NA60 data from an independent analysis~\cite{Shahoyan:2006qm} of the
intermediate mass region (``IMR'') 1.16$<$$M$$<$2.56 GeV, corrected
for the contribution from Drell-Yan pairs, are shown for
comparison. Finally, the hadron data obtained as a by-product of the
cocktail subtraction procedure are also included; the value for the
$\eta$ has been obtained by tuning the GENESIS
code~\cite{genesis:2003} to the $\eta$ Dalitz decay and then referring
back to the required $T_\mathrm{eff}$ of the mother. The errors shown
for the low-mass data (``LMR'') are purely statistical. Systematic
errors only enter to the extent that the slopes of the $p_{T}$ spectra
are affected, not their absolute level. The errors due to the
background subtraction are about the same as the statistical errors of
the fine-bin data ($\sim$7 MeV). The errors associated with the
subtraction of the decay sources, though significant for the yields
(see above), lead to errors of only 1-4 MeV for $T_\mathrm{eff}$
(dependent on the source), due to both the mostly local nature of the
subtraction and the closeness of $T_\mathrm{eff}$ for dimuons and
hadrons (see Fig.~\ref{fig4}). All other error sources considered -
relative acceptance, the sensitivity to the input $y$-distribution,
cuts vs. no cuts in $y$ (none are used), cuts vs. no cuts in
cos$\theta_{CS}$ (none are used) - also lead to differences
considerably smaller than the statistical errors. A correction for
Drell-Yan, using an extrapolation down to $M$$<$1
GeV~\cite{vanHees:2006ng}, would systematically lower the values by
5-10 MeV, depending on mass.

The results displayed in Fig.~\ref{fig4} can be summarized and
interpreted as follows. The slope parameter $T_\mathrm{eff}$ rises
nearly linearly with mass up to about 270 MeV at the pole position of
the $\rho$, followed by a sudden decline to values of 190-200 MeV for
masses $>$1 GeV. The excess yield in the mass region 0.2$<$$M$$<$0.9
GeV has generally been attributed to thermal radiation from the
fireball, dominated by pion annihilation
$\pi^{+}\pi^{-}\rightarrow\rho\rightarrow\mu^{+}\mu^{-}$, and the NA60
data, before acceptance correction, have directly been interpreted as
the space-time averaged in-medium spectral function of the
$\rho$~\cite{Arnaldi:2006jq}. Following earlier
work~\cite{Rapp:1999ej}, they are now nearly quantitatively described
by the newest theoretical
developments~\cite{vanHees:2006ng,Ruppert:2006hf,Dusling:2007rh}. The
linear rise of $T_\mathrm{eff}$ with $M$ over the whole region up to
the $\rho$ peak is reminiscent of radial flow of a hadronic
source. 
While the hadron data show a similar linear rise, their
absolute values are surprisingly close to the excess values, contrary
to the expectation for the temperature-flow folding expressed in the
introduction. The seeming contradiction can be resolved by comparing
the (free) $\rho$ itself rather than the other hadrons with the
in-medium emission. This $\rho$ is accessible as the peak on the broad
continuum (Fig.~\ref{fig1}), generally interpreted as the freeze-out
$\rho$ without in-medium
effects~\cite{vanHees:2006ng,Ruppert:2006hf,Dusling:2007rh}. By
disentangling the peak from the continuum as done
before~\cite{Damjanovic:2006bd,Damjanovic:2007qm}, we find
$T_\mathrm{eff}$=300$\pm$17 MeV for the peak and 231$\pm$7 for the
underlying continuum in the window 0.6$<$$M$$<$0.9 MeV. The high value
of the peak, added as a further hadron point into Fig.~\ref{fig4},
should then be interpreted as characteristic for the true freeze-out
parameters of the fireball, implying the $\eta$, $\omega$ and $\phi$
to freeze out earlier, due to their smaller coupling to the pions. By
modeling a $\rho$ with this temperature, its contribution can be
subtracted from the total for each of the finer binned data points in
Fig.~\ref{fig4}, lowering $T_\mathrm{eff}$ by 4$-$20 MeV depending on
the closeness to the pole, and shifting the maximum of the resulting
``in-medium'' values up to the mass bin just below 1 GeV. In any case,
the large gap in $T_\mathrm{eff}$ between the vacuum $\rho$ and the
excess points (corrected or not for the subtraction) restores the
expected difference between a freely emitted hadron and its in-medium
decay part, making the observed linear rise of $T_\mathrm{eff}$ with
$M$ now {\it consistent} with the expectations for radial flow of a
{\it hadronic} source (here $\pi\pi \rightarrow \rho$) decaying into
lepton pairs. Theoretical modeling of our results is
underway~\cite{vanHees:2006ng,Ruppert:2006hf,Dusling:2007rh}, but does
not yet describe the data in a satisfactory way.

It is interesting to note that the large gap of $>$50 MeV in $T_\mathrm{eff}$
between the vacuum $\rho$ and the $\omega$ (same mass) decreases
towards the peripheral window, but only closes for the lowest
peripheral ``pp-like'' selection 4$<$$dN_{ch}/d\eta$$<$10 shown in
Fig.~\ref{fig3} (lower), with $T_\mathrm{eff}$=198$\pm$6 MeV for the $\rho$
and 201$\pm$4 MeV for the $\omega$. This implies that both the ``hot
$\rho$'' and the low-$m_{T}$ rise discussed before are intimately
connected to pions, disappearing together as the $\pi\pi$ contribution
to $\rho$ production vanishes (with only the cocktail $\rho$ left).

The sudden decline of $T_\mathrm{eff}$ at masses $>$1 GeV is the {\it other}
most remarkable feature of the present data. Extrapolating the
lower-mass trend to beyond 1 GeV, a jump by about 50 MeV down to a
low-flow situation is extremely hard to reconcile with emission
sources which continue to be of dominantly hadronic origin in this
region. {\it If} the rise is due to flow, the sudden loss of flow is
most naturally explained as a transition to a qualitatively different
source, implying dominantly early, i.e. {\it partonic} processes like
$q\bar{q}\rightarrow\mu^{+}\mu^{-}$ for which flow has not yet built
up~\cite{Ruppert:2006hf}. While still
controversial~\cite{vanHees:2006ng}, this may well
represent the first direct evidence for thermal radiation of partonic
origin, breaking parton-hadron duality for the {\it yield} description
in the mass domain.

In conclusion, we have found strong evidence for radial flow in the
region of thermal dilepton emission which has previously been
associated with the $\rho$ spectral function. The transition to a
low-flow region above may signal a transition from a hadronic to a
partonic source.


\begin{acknowledgments}
We are grateful to H.~van~Hees, R.~Rapp, T.~Renk and J.~Ruppert for
useful discussions. We acknowledge support from the BMBF (Heidelberg
group) as well as the C. Gulbenkian Foundation, Lisbon, and the Swiss
Fund Kidagan (YerPHI group).
\end{acknowledgments}

\end{document}